\documentclass[a4paper,11pt]{article}
\usepackage{amsmath}
\usepackage{a4wide}
\usepackage{pstricks,pst-node,pst-text,pst-3d}
\usepackage{amsfonts}
\usepackage{amssymb}
\usepackage{graphicx,graphics,epsfig}
\usepackage{geometry}
\usepackage[english]{babel}
\usepackage[utf8]{inputenc}
\usepackage[T1]{fontenc}
 \usepackage{float}
\usepackage{hyperref}
\usepackage{cite}

\numberwithin{equation}{section} 
\def\noi{\noindent}
\def\non{\nonumber}

\newcommand{\Lag}{\mathcal{L}}

\newcommand{\chibar}{\bar{\chi}}
\newcommand{\mchi}{{m_{\chi}}}
\newcommand{\neuti}{{\tilde{\chi}}_i^0}
\newcommand{\neuto}{{\tilde{\chi}}_1^0}
\newcommand{\mneuto}{m_{{\tilde{\chi}}_1^0}}
\newcommand{\chargi}{{\tilde{\chi}}_i^\pm}
\newcommand{\chargj}{{\tilde{\chi}}_j^\mp}
\newcommand{\chargopm}{{\tilde{\chi}}_1^\pm}

\newcommand{\SloopS}{{\texttt{SloopS}}}

\newcommand{\Lanhep}{\texttt{LanHEP\,}}
\newcommand{\FormCalc}{{\texttt{FormCalc}}}

\newcommand{\LoopTools}{\texttt{LoopTools}}

\newcommand{\tb}{t_\beta}
\newcommand{\cosb}{c_\beta}
\newcommand{\sinb}{s_\beta}
\newcommand{\higgzino}{{\tilde{H}}}
\newcommand{\singino}{{\tilde S}}

\newcommand{\wino}{\tilde{W}}
\newcommand{\bino}{\tilde{B}}

\newcommand{\sigmav}{\langle \sigma v \rangle}

\newcommand{\ra}{\rightarrow}

\begin{document}
\begin{titlepage}
 \vspace*{0.1cm}
 \rightline{TTP11-28}
 \rightline{LAPTH-038/11}

\begin{center}
 {\bf Loop-induced photon spectral lines from neutralino annihilation in the
NMSSM.}

\vspace{0.5cm}
{G.~Chalons${}^{1)}$, A.~Semenov${}^{2,3)}$}\\

\vspace{4mm}
{\it 1) Institut f\"ur Theoretische Teilchenphysik, \\
Karlsruhe Institute of Technology, Universit\"at Karlsruhe \\
Engesserstraße 7, 76128 Karlsruhe, Germany} \\ \vspace{4mm}
{\it 2) Laboratory of Particle Physics, Joint Institute for Nuclear Research,\\
141 980 Dubna, Moscow Region, Russian Federation }\\ \vspace{4mm}
{\it 3) LAPTH, Universit\'e de Savoie, CNRS, \\
BP 110, F-74941 Annecy-le-Vieux Cedex, France}

\vspace{10mm}

\abstract{
We have computed the loop-induced processes of neutralino annihilation into two
photons and, for the first time, into a photon and a $Z^0$ boson in the
framework of the NMSSM. The photons produced from these radiative modes are 
monochromatic and possess a clear ``smoking gun'' experimental signature. This
numerical analysis has been done with the help of the \SloopS~code, initially
developed for automatic one-loop calculation in the MSSM. We have computed the
rates for different benchmark points coming from SUGRA and GMSB soft SUSY
breaking scenarios and compared them with the MSSM. We comment on how this
signal can be enhanced, with respect to the MSSM, especially in the low mass
region of the neutralino. We also discuss the possibility of this observable to
constrain the NMSSM parameter space, taking into account the latest limits from
the \texttt{FERMI} collaboration on these two modes.
}
\end{center}
\normalsize
\end{titlepage}

\section*{Introduction}
The existence of cold dark matter (CDM) is supported by many geometrical and
dynamical observations coming from cosmology and astrophysics, but its
detection needs to be confirmed. Its properties, such as mass, spin,
and interactions are still to be determined. A long term experimental effort
has been devoted to its detection, from direct methods, where a dark matter
particle is expected to impinge on a nuclei in a target material, or more
generally from the products of its self-annihilation in outer space (indirect
detection).
\par\noi
Recently the \texttt{CDMS} \cite{cdmsII10},
\texttt{CoGeNT} \cite{cogent10,cogent11} experiments and very recently
\texttt{CRESST} \cite{cresst11} have pointed out
results whose explanation could be interpreted as hints in favour of the
existence of low mass dark matter particles, although the
recent result of the \texttt{XENON100} disfavors this
possibility \cite{xenon100-11} and the \texttt{CDMS-II} results seems to be in
contention with \texttt{CoGeNT} \cite{cdmsII10}. The interesting piece of
information is that these results could be reconciled with the long standing
claims of the \texttt{DAMA/LIBRA} experiment that dark matter was detected
through its annual modulation \cite{dama10}. However these potential signals
of dark matter are still an open debate and no firm claim of dark matter
signal can be established, since there is no general agreement between the
different experiments. If a dark matter hypothesis is to be put forward to
interpret these signals, the studies points to a low mass WIMP. Thus some
attempts were made to account for them, for example with neutralinos
lighter than, say, 15 GeV. In recent publications the authors of
\cite{daniel-nmssm, daniel-mssm,das10,draper10,kappl10,cao11} considered and
investigated the occurrence of light neutralino within the \textit{Minimal
Supersymmetric Standard Model} (MSSM) and  {\it Next-to-Minimal Supersymmetric
Standard Model} (NMSSM) while respecting several constraints. It was shown that
an MSSM explanation for the signal is seriously challenged by collider searches
and measurements in the flavour sector
\cite{daniel-nmssm,daniel-mssm,fornengo11,calibbi11}. 
\par\noi 
The {\it Next-to-Minimal Supersymmetric Standard Model} is a
well-motivated extension of the MSSM constructed by adding a singlet superfield
$\hat S$ in the Higgs sector. This has been advocated to provide an elegant
solution to the so-called $\mu$ problem of the MSSM but it also renders the
Higgs ``little fine tuning problem'' less severe \cite{nmssmrev-ellwanger}.
Moreover it leads to a richer spectrum than the MSSM : two additional Higgs
bosons (one CP-even and one CP-odd neutral Higgs) remain after ElectroWeak
Symmetry Breaking (EWSB) and the neutralino sector is enlarged by a fifth
neutralino. This last possibility has important consequences concerning dark
matter studies, that can differ significantly from the MSSM, especially when the
lightest neutralino composition is dominated by the singlet component, the
singlino. Moreover the occurrence of very-light singlet pseudoscalar in the
Higgs spectrum and at the same time evading the LEP bounds is still possible
thanks to reduced couplings to standard model particles. These facts open new
possibilities in relic density calculation as well as direct detection prospects
as opposed to the MSSM. In particular the thermally produced neutralinos can
satisfy the measured CDM present relic abundance while being very light. This
implies that the pair annihilation rate of thermal relics roughly scales as
$1/\mchi^2$. All these observations make a light supersymmetric candidate for
dark matter more plausible in the NMSSM than the MSSM. It is then appealing to
investigate the self annihilation of dark matter in our galaxy since one expects
light neutralinos to give significantly enhanced rates for the indirect
detection signals compared to the standard case.
\par\noi 
Numerous studies have been devoted to relic density and direct
detection phenomenology in the NMSSM, few work has been dedicated to indirect
detection of dark matter (annihilation of a pair of dark matter particle)
through the ``direct'' annihilation into primary monochromatic photons
\cite{profumogg}. These primaries photons have the advantage of being less
affected by astrophysical uncertainties over other kind of messengers. Moreover
their spectrum would reveal a sharp peak at an energy $E_\gamma \simeq M_\chi$
corresponding to the mass of the dark matter
particle for the $\gamma \gamma$ final state and $E_\gamma \simeq M_\chi (1 -
M_Z^2/4 M_\chi^2)$ for $\gamma Z^0$, since in the galactic halo $v/c \simeq
10^{-3}$. Provided one has a good enough detector energy resolution and
sensitivity, the flux from these primaries photons will be clearly distinctive
from the astrophysical background or diffuse emission. However modeling the
dark matter halo is still needed since the number
density of dark matter particles enters the calculation. The \texttt{FERMI}
collaboration did not report any line observation and has released upper limits
on the direct annihilation $\sigmav_{\gamma\gamma}$ and $\sigmav_{\gamma Z}$
\cite{fermigg} instead. The recently installed \texttt{AMS} \cite{ams} detector
on the International Space Station may also shed light on these channels. 
\par\noi 
The complete computation of the loop-induced annihilation into photons has
already been performed in \cite{lsptogg,lsptozg,lsptoggfernand,boudjema05} for
the MSSM. In \cite{profumogg} the
one-loop amplitudes for NMSSM neutralino pair annihilation into two photons and
two gluons have been given (adapted to the NMSSM case from the formulas given
in \cite{lsptogg}) and prospects for the indirect detection of the
monochromatic gamma-ray line were also investigated. In light of the latest
limits of the \texttt{FERMI} collaboration on monochromatic gamma ray signal and
the recent activity in the dark matter community, we
propose to revisit the two gammas mode and provide for the first time results on
the $\gamma Z^0$ one. We investigated to which extent the NMSSM rates could 
differ from the MSSM and for the numerical study we focused on mSUGRA and GMSB
benchmarks points given in \cite{bmp-nmssm,ellwanger-gmsb}. This is a case study
if one wants to discriminate the NMSSM from the MSSM with dark matter related
observables, which basically means to which amount the neutralino sector is
sensitive to the extended Higgs sector.  We further explored in which case,
specific to the NMSSM, the spectral line can be enhanced and if such mechanisms
can be constrained by astrophysical measurements. Constraining these
mechanisms is very interesting since it also impacts the relic density
calculation and direct detection predictions.
\par\noi
 These rates were computed with the help of the \SloopS~program, an automatic
code for one-loop calculation in the context of the SM and the MSSM
\cite{baro08,baro09}. This code has already been used for accurate relic
density predictions at next-to-leading order in the MSSM
\cite{baro07,barosusy09,chalons09} and also to the numerical computation of the
indirect channels $\neuto\neuto \ra \gamma\gamma, \gamma Z^0$ and two gluons
mode \cite{boudjema05}. Therefore this work is also a good exercise to test the
implementation of the NMSSM in \SloopS. On the technical level, as the dark
matter particles are moving at relatively small velocity, the Gram determinant,
which is a key ingredient for calculating the loop integrals, vanishes and this
results in numerical instabilities. This has been handled with a procedure
called segmentation \cite{boudjema05}. We will come back to the numerical and
technical details of the implementation later. 
The outline of the paper is as follows, in section one we quickly review the
NMSSM model and its parameter space, in section two we describe the \SloopS~code
and the implementation of the NMSSM. A particular attention will be paid to
the implementation of non-linear gauge-fixing in the NMSSM. The third section
will be devoted to scrutinise the additional contributions brought when going
from the MSSM to the NMSSM. In the fourth section we will give the rate
for the $\gamma\gamma$ and $\gamma Z^0$ channels for some SUGRA and GMSB
scenarios and compare it to an equivalent MSSM spectrum. In the following
section we will discuss on some possible ways of increasing the signal, specific
to the NMSSM, and discuss the relevance of using the sharp gamma lines as a
constraining observable, taking into account the latest published limits.
Finally we will draw our conclusions.
\section{Overview of the NMSSM}
In the NMSSM the Higgs term of the superpotential involving the two Higgs
doublet is modified  and a singlet term is added\footnote{We stick to the
``$\mathbb{Z}_3$-invariant NMSSM'',
where any dimensionful parameters in the superpotential are
forbidden.}\cite{nmssmrev-ellwanger},
\begin{equation}
 W_{NMSSM} = W_{MSSM}^{\mu=0} + \lambda \hat S \hat H_u \hat H_d +
\frac{\kappa}{3}\hat S^3
\end{equation}\noi 
The MSSM $\mu$ bilinear term is now absent from the superpotential and has been
replaced by the trilinear coupling of the singlet with the two Higgs doublets.
The VEV of the singlet generates an effective $\mu$ parameter with respect to
the MSSM, which is then naturally of order the EW scale
\cite{nmssmrev-ellwanger},
\begin{equation}
 \mu_{\mathrm{eff}} = \lambda s
\end{equation}\noi 
where $s = \langle \hat S \rangle$ is the VEV of the Higgs singlet. The
soft-SUSY breaking Lagrangian is also modified according to
\begin{eqnarray}
 -\Lag_{\mathrm{soft}} &=& m^2_{H_u} |H_u|^2 +  m^2_{H_d} |H_d|^2 + m_S^2 |
S|^2\non \\
&+&(\lambda A_\lambda H_u \cdot H_d S + \frac{1}{3} \kappa A_{\kappa} S^3 + h.c)
\end{eqnarray}\noi 
Given $M_Z$ and using conditions coming from the minimisation of the Higgs
potential, one can choose six independent parameters for the Higgs sector 
\begin{equation}
 \lambda, \kappa, A_\lambda, A_\kappa, \mu_{\mathrm{eff}}, \tb
\end{equation}\noi 
where $\tb = \mathrm{tan} \beta = v_u/v_d$, the ratio of the two Higgs doublet
VEV's : $\langle H_u^0 \rangle = v_u,\langle H_d^0 \rangle = v_d$ . This is in
contrast with the MSSM where two parameters are needed, e.g, $\mathrm{tan}
\beta$ and the mass of the Higgs pseudoscalar $M_{A^0}$, once the requirement of
vanishing tadpoles has been imposed. After EWSB the NMSSM Higgs
sector contains three neutral scalar fields, $H_1,H_2,H_3$ and two pseudoscalar
neutral ones, $A_1, A_2$ as well as a charged Higgs $H^{\pm}$. In the
neutralino sector the additional singlino mixes with the bino, wino and
Higgsinos fields. The neutralino mass matrix is therefore a $5\times 5$ one
which is diagonalised with a unitary matrix $N$. In the basis $\tilde\chi^0 = (
-i\tilde B, -i\tilde W_3, \higgzino_u^0, \higgzino_d^0, \tilde S^0)$ the
neutralino mass matrix reads, 
\begin{equation}
 \begin{pmatrix}
  M_1 &      0 & -\cosb s_W M_Z & \sinb s_W M_Z            & 0 \\
   0     & M_2 & \cosb c_W M_Z  & - \sinb c_W M_Z          & 0 \\
   -\cosb s_W M_Z & \cosb c_W M_Z & 0   &-\mu_{\mathrm{eff}}&-\lambda v_u\\
   \sinb s_W M_Z &   - \sinb c_W M_Z   &-\mu_{\mathrm{eff}} & 0  &-\lambda v_d\\
    0    &   0     &  -\lambda v_u\ & -\lambda v_d&   2 \kappa s\\
 \end{pmatrix}
\end{equation}\noi 
The lightest supersymmetric particle (LSP) can then be expressed as a linear
combination of the five gauge eigenstates
\begin{equation}
 \neuto = N_{11} \bino + N_{12} \wino + N_{13} \higgzino_1 + N_{14} \higgzino_2
+ N_{15} \singino
\end{equation}\noi 
For a pure state the singlino mass is
\begin{equation}
 m_{\singino} = 2 \kappa s
\end{equation}\noi
The parameter space of the NMSSM can then be described, in addition to the six
Higgs sector parameters, by the same as the MSSM, namely the soft masses for
sfermions $M_{\tilde f}$, trilinear couplings $A_f$ and gaugino masses
$M_{1,2,3}$ which are respectively the $U(1)$, $SU(2)$ and $SU(3)$ soft
parameters. If the LSP has a dominant singlino component, it can efficiently
annihilate through light singlet Higgses as well as light pseudoscalar Higgs
singlet \cite{profumogg,belanger-nmssm}. The MSSM limit is recovered when
$\lambda, \kappa \ra 0$ and the equivalent of the mass (squared) of the only
physical CP-odd scalar $A^0$ of the MSSM is given by
\begin{equation}
\label{MA0MSSM}
 M_{A^0}^2 = \frac{2 \lambda s (A_\lambda + \kappa s)}{\mathrm{sin} 2\beta}
\end{equation}\noi 
This is an important proviso when we will compare the NMSSM result with an
''equivalent`` MSSM spectrum.

\section{Set-up of the automatic calculation}
One loop processes calculated via the diagrammatic Feynman approach imply the
calculation of hundreds of Feynman diagrams and a hand calculation is proned to
numerous errors. A high-level of automation is therefore highly desirable,
especially if one wants to build a general purpose code. The \SloopS~code
\cite{boudjema05,baro08,baro09} has been developed in this purpose and applied
to astrophysics \cite{boudjema05}, cosmology \cite{baro07,barosusy09,chalons09}
and also collider physics \cite{baro09,Hao09}, both in the SM and MSSM. The
implementation of the NMSSM has been carried out with \Lanhep \cite{lanhep}
which generates the complete set of NMSSM vertices once the Lagrangian is
specified. The calculation of the process $\neuto\neuto \ra \gamma Z^0$
requires the field renormalisation $\delta Z^{1/2}_{Z\gamma}$ which is generated
from the (tree-level) $\neuti\neuto Z$ vertex through a $Z-\gamma$ one-loop
transition. This renormalisation constant is generated in \Lanhep by shifting
the appropriate fields and defined in the on-shell scheme. The output files are
then written in the \FormCalc \cite{formcalc} conventions which handle the
calculation of the cross section. One nice feature of \SloopS~is the use of a
generalised non-linear gauge fixing \cite{chopin-nlg}, adapted to the
supersymmetric case \cite{boudjema05,baro08}, and for this particular work we
have extended it to the specific NMSSM Higgs sector. Compared to the MSSM, the
gauge-fixing Lagrangian will depend on two additional parameters since the
Higgs sector is enlarged with one more scalar and one more pseudoscalar. This
allows us to perform a non-trivial check of the gauge independence of the result
through the variation of eleven non-linear gauge parameters. The gauge fixing
Lagrangian reads in a general form
\begin{equation}
\label{gaugefixing}
{\mathcal L}_{GF} = -\frac{1}{\xi_W} F^+ F^- - \frac{1}{2  {{ \xi_Z}}  }|
F^Z|^2 - \frac{1}{2 {{ \xi_A}} } | F^A|^2
\end{equation}\noi 
where the non-linear functions of the fields $F$ are given by
\begin{eqnarray}
F^+ & = & \bigg(\partial_\mu - ie {\tilde{\alpha}}  A_\mu - igc_W 
{\tilde{\beta}} Z_\mu\bigg) W^{\mu \, +}\non \\
&  &+ i{{ \xi_W}}  \frac{g}{2}\bigg(v +  {\tilde{\delta}}_1
H_1 + {\tilde{\delta}_2} H_2  +{\tilde{\delta}_3} H_3 + i({\tilde{\kappa}} G^0+
{\tilde{\rho}}_1 A_1 + {\tilde{\rho}}_2 A_2)\bigg)G^+\\
F^Z & = & \partial_\mu Z^\mu + {{ \xi_Z}}  \frac{g}{2c_W}\bigg(v +
    {\tilde{\epsilon}}_1  H_1 +  {\tilde{\epsilon}}_2 H_2 +  
{\tilde{\epsilon}}_3 H_3  \bigg)G^0 \\
F^A & = & \partial_\mu A^\mu
\end{eqnarray}\noi 
where $G^0$ and $G^\pm$ are respectively the neutral and charged goldstones.
The parameters $\tilde \alpha,\tilde \beta \cdots \tilde \epsilon_3$ are
generalised gauge fixing parameters.
\par\noi
The ghost Lagrangian $\mathcal{L}^{Gh}$ is derived by requiring that the full
effective Lagrangian is invariant under BRST transformations. This implies 
that the full quantum Lagrangian, with $\Lag_C$ and
$\Lag_{Gh}$ respectively the classical and ghost Lagrangians,
\begin{equation}
 \Lag_Q  = \Lag_C + \Lag_{GF}+ \Lag_{Gh}
\end{equation}\noi 
be such that $\delta_{\mbox{\tiny BRS}} \Lag_Q = 0$ and hence
$\delta_{\mbox{\tiny BRS}}\Lag_{GF} = - \delta_{\mbox{\tiny BRS}}\Lag_{Gh}$
\cite{grace-1loop}. The BRST transformation for the gauge fields can be found
for example in \cite{grace-1loop}. The NMSSM specific transformations for the
scalar fields can be found in the Appendix. Within this particular gauge fixing
we can set $\xi_{W,Z,A} = 1$ (avoiding complicated tensor structure for the
gauge bosons propagators) and keep the possibility to check the gauge invariance
of the result, at the expense of adding new vertices to the model.
\par\noi
The version of \LoopTools \cite{looptools} we used is a modified one which
tackles the problem of inverse Gram determinant.  This kinematic quantity
vanishes when the velocity $v$ is equal to zero. Consequently the reduction
algorithm expressing the tensorial loop integrals onto a basis of scalar ones
breaks down. The details of this procedure can be found in \cite{boudjema05}. In
short this method uses the particular kinematics at $v=0$ to reduce, for
example, scalar box integrals to a sum of triangle ones. The neutralino mass
matrix is diagonalised numerically with a complex unitary matrix $N$ giving
positive eigenvalues for the neutralino masses.
\section{Additional NMSSM contributions}
In the MSSM, depending on the nature of the neutralino, the predicted rates
for $\sigmav_{\gamma\gamma/Z}$ can be quite different. The highest ones are
reached when the LSP is mostly wino or higgsino like. In the former case the
dominant contributions are from loops with gauge bosons, since in the wino case
the most important coupling is $\neuto\chargopm W^\mp$. For a Higgsino $\neuto$
the dominant contribution would be vertex corrections through a $Z^0$ boson in
the s-channel as the $\neuto\neuto Z$ coupling is proportional to 
\begin{equation}
 g_{\neuto\neuto Z} \propto N_{13}^2 - N_{14}^2
\end{equation}\noi
The third possible case, a bino-like neutralino, would annihilate mainly through
box diagrams containing right-handed sfermions since the bino, as the
superpartner of the $B$ boson, couples to sparticles possessing the highest
hypercharge. A bino like LSP gives generically lower rates than the two other
cases as its couplings to other particles are weaker. Therefore if one wants to
stick to a mostly bino neutralino while giving good prospects for the spectral
photon lines, significant mixing with the other components is needed.
\par
\begin{figure*}[t]

\begin{center}
\epsfig{file=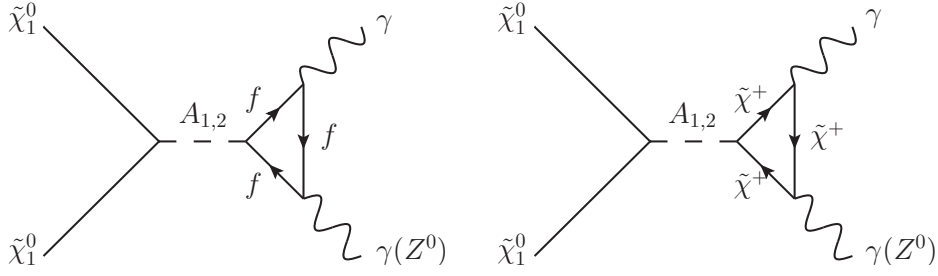,clip=,width=0.7\textwidth}
\caption{\label{NMSSMdiags} \em Additional NMSSM diagrams with s-channel
pseudoscalar exchange for the processes $\neuto\neuto \ra \gamma \gamma
(Z^0)$. The label $f$ stands for a SM fermion and $\chi^+$ to a chargino.} 
\end{center}
\end{figure*}

\noi 
In going from the MSSM calculation to the NMSSM one, two major differences
appear : on the one hand the NMSSM couplings now depend on the parameters
$\lambda$ and $\kappa$, and on the other hand new diagrams are to be computed.
Concerning the case of the modified couplings, if the NMSSM neutralino is
a mostly bino, wino or higgsino, we do not expect major differences with the
MSSM, as long as the singlino component is negligible.
As the neutralino is a Majorana fermion, at vanishing relative velocity the LSP
pair carries pseudoscalar quantum numbers. Therefore the NMSSM additional
diagrams are the ones with pseudoscalar $A_1,A_2$ s-channel exchange. The
relevant diagrams are depicted in Fig.~\ref{NMSSMdiags}. Note that the diagram
with the chargino loop should be understood as containing both ''flavours`` of
charginos since the couplings $A_{k}\chargi \chargj$ are non-diagonal. Note
also that the couplings $A_{k} \tilde f_i \tilde f_j$ are also non-diagonal in
family space because we stick to real soft SUSY breaking terms. Hence one could
think that the NMSSM brings also new contributions containing loops with
different flavours of sfermions. In fact their amplitudes cancel when summing
over all the flavours\footnote{This is of course also the case in the MSSM.}.
Furthermore due to our particular gauge-fixing Lagrangian in
Eq.~(\ref{gaugefixing}), we generate couplings of the pseudoscalars to the
charged goldstones $G^\pm$ and to the charged ghosts $c^W, \bar c^W$,
proportional to $\tilde \alpha$ and $\tilde \rho_{1,2}$. These additional
couplings are not present in the linear gauge. However, just like the case of
the sfermions coupling to $A_{1,2}$ the related amplitudes vanish identically.
This is however not surprising since there exists no ''tree-level`` coupling of
the pseudoscalars to a pair of gauge bosons. In turn, the process with $\gamma
\gamma$ will involve only the non-linear gauge parameter $\tilde\alpha$ and the
process with $\gamma Z^0$ the parameter $\tilde\beta$ in addition. Obviously the
final result should not depend on these two parameters, they are used as a check
on the gauge invariance of the cross sections, which we ascertained.
\par\noi 
The use of the non-linear gauge fixing Lagrangian of Eq.~(\ref{gaugefixing}) in
the 't Hooft-Feynman gauge ($\xi_A =\xi_Z = \xi_W = 1$) also enables us to
cancel, for example, the $W^+ G^-  \gamma$ vertex, by setting $\tilde\alpha=-1$,
which cancels a corresponding piece of the original trilinear sector of the
Lagrangian, leading to a vanishing total $W^+ G^- \gamma$ trilinear coupling
(see \cite{bergstrom94gg} and references therein).  We can perform the same
trick for the coupling $W^+ G^-  Z^0$ by setting $\tilde \beta$ to an
appropriate value.  Finally, as a further check on the
implementation we took the input parameters for the six scenarios presented in
\cite{boudjema05}, compared with our results by taking the MSSM
limit, and found a perfect agreement.

\par\noi
The \texttt{FERMI} collaboration has already provided their limits
\cite{fermigg} on the $\gamma \gamma$ and $\gamma Z^0$ channels for different
photon energies $E_{\gamma}$, hence we did not model the propagation of
the
photon signal and we restricted ourselves to the discussion of
$\sigmav_{\gamma\gamma}$ and $\sigmav_{\gamma Z}$. 
\section{Benchmark points and comparison with the MSSM}
As a first application of our code we computed the gamma-ray lines for the
SUGRA and GMSB benchmark points from \cite{bmp-nmssm,ellwanger-gmsb} and
\texttt{NMSSMTools\_2.3.5} \cite{nmssmtools}. For a discussion about these
benchmarks we refer to their respective publication
\cite{bmp-nmssm,ellwanger-gmsb}. We then compared with the present limit given
by \texttt{FERMI} \cite{fermigg} and against the MSSM calculation, performed
with the \texttt{SloopS} code with a MSSM model file, like in \cite{boudjema05},
to see how much precision one would need to distinguish them. The input
parameters for the MSSM parameters are the same as in the NMSSM taken in the
decoupling limit $\lambda,\kappa \ra 0$ and using Eq.~(\ref{MA0MSSM}). It is
important to note that, in order to preserve gauge invariance, we only took the
input parameters obtained after the RGE running (therefore $\overline{\rm DR}$
parameters) to the low-energy scale and considered them as \textit{physical}
parameters, no mass correction or effective couplings have been considered. The
self-annihilation of the neutralino is mainly driven by its mass and
composition, so the first question to ask when comparing the MSSM with the NMSSM
is whether their mass and composition are similar. 

\subsection{SUGRA}
The SUGRA input parameters are taken from the version \texttt{2.3.5}
of \texttt{NMSSMTools} as they have been substantially modified since the
publication of \cite{bmp-nmssm}. The most relevant parameters are given in
Table~\ref{inputSUGRA}.
\begin{table}[htbp]
 \begin{center}
  \begin{tabular}{|c|c|c|c|c|}
   \hline
   Parameter & SUGRA-P1 & SUGRA-P2 & SUGRA-P3 & SUGRA-P4 \\
   \hline
   $M_1$ & 211.62&211.62 &211.86 & 333.80\\
   $M_2$ & 391.86&391.86 &392.51 & 613.91\\
   $\mu$ & 968.62 &968.62 &938.41 & -206.17\\
   $\tb$ & 10.000& 10.000&10.000 & 3.0413\\
   $ M_{\tilde l_{1,2}^R}$ & 257.51&257.51 &256.38 &702.15\\
   $ M_{\tilde l_3^R}$ & 221.13&221.13 &221.21 &698.94\\   
   $ M_{\tilde u_{1,2}^R}$ &986.12 &986.12 &985.64 &1674.5\\
   $ M_{\tilde u_3^R}$ &565.23 &565.23 &590.90 &527.83\\  
   $ M_{\tilde d_{1,2}^R}$ & 981.60&981.60 &981.17 &1609.0\\
   $ M_{\tilde d_3^R}$ & 966.15&966.15 &966.20 &1606.6\\
   $A_\mu$ &-1777.5 &-1777.5 &-1698.7 & -2531.2\\
   $A_\tau$ & -1763.0& -1763.0&-1682.8 & -2524.8\\
   $A_b$ &-2611.8 & -2611.8&-2535.2 & -3730.5\\
   $A_t$ &-1431.5 & -1431.5 &-1362.6 & -1765.8\\
   \hline
   \multicolumn{5}{|c|}{MSSM specific parameter} \\
  \hline
   $M_{A^0}$   &943.65 &943.65 &935.19 & 702.11 \\
   \hline
   \multicolumn{5}{|c|}{NMSSM specific parameters} \\
   \hline
   $\lambda$ &0.1 & 0.1& 0.4&  0.49\\
   $\kappa$ &0.1089 & 0.1089& 0.3037& 0.0554 \\
   $A_\lambda$ & -963.91& -963.91& -620.23& -684.51\\
   $A_\kappa $ & -1.5893& -1.0934& -11.158& 151.73\\
   \hline 
  \end{tabular}
\caption{{\em SUSY input parameters for the SUGRA benchmarks. Masses are in
GeV.}\label{inputSUGRA}}
 \end{center}
\end{table}\noi
The first three scenarios (P1, P2, P3) exhibit a mostly bino-like neutralino and
are quite similar, as far as only the neutralino/chargino sector is concerned.
In the four cases the lightest neutralino is heavy enough ($\mneuto \simeq 211$
GeV) such that the $\gamma Z^0$ final state is open. As the first three
benchmark scenarios give a bino-like LSP, we can already anticipate that the
NMSSM and MSSM will give similar results as they have a very similar
composition. However we expect that the fourth scenario P4 will give a clearly
distinctive signature since in this case the lightest NMSSM neutralino $\neuto$
is 99\% singlino with mass $\mneuto \simeq 60$ GeV. The final
results can be found in Table~\ref{resSUGRA}. 
\begin{table}[htbp]
 \begin{center}
  \begin{tabular}{|c|c|c||c|}
 \hline 
 Model & $\mchi$ (GeV)& $\sigmav_{\gamma\gamma}\times
10^{30}[\mbox{cm}^3\, \mbox{s}^{-1}]$ & $\sigmav_{\gamma Z}\times
10^{30}[\mbox{cm}^3\, \mbox{s}^{-1}]$ \\
\hline
\multicolumn{4}{|c|}{SUGRA-P1/P2}  \\
\hline
NMSSM & 210.74 &$3.9840$ & $1.7391$\\
MSSM   & 210.74 & $3.9836$ &$1.7393$ \\
\hline
\multicolumn{4}{|c|}{SUGRA-P3}  \\
\hline
NMSSM & 210.92 &$4.0103$& $1.7326$\\
MSSM   & 210.93 & $4.0074$ & $1.7348$\\
\hline 
\multicolumn{4}{|c|}{SUGRA-P4} \\
\hline
NMSSM & 59.963 &$2.7157\,10^{-3} $ & $2.1012\,10^{-3}$\\
MSSM   & 200.43 & $61.866$ & $205.28$\\
\hline
\end{tabular}
\caption{{\em Rates for the loop-induced annihilation of two
neutralinos into $\gamma\gamma/Z^0$ for the SUGRA benchmarks.}\label{resSUGRA}}
 \end{center}
\end{table}
\par\noi 
We see that, unless the NMSSM neutralino has a radically different
composition from the MSSM case, these two frameworks are almost
indistinguishable. Needless to say that this is also a good indication of the
implementation of the NMSSM model since in the decoupling limit (which is the
case for these two scenarios) we should recover the MSSM. 
The benchmarks SUGRA-P1 and SUGRA-P2 give equal results because they only
differ by the value of $A_\kappa$, which has an impact only on the mass of the
lightest pseudoscalar, which is not relevant in these cases. A difference is
clearly visible when the NMSSM neutralino is mostly singlino,
see the SUGRA-P4 case, whereas the MSSM neutralino is higgsino-like. Obviously
for this latter case the comparison is not very meaningful from an experimental
point of view since the inputs would be the energy of the photon $E_\gamma$ and
the rate $\sigmav$, which are clearly different in this benchmark point.
Considering only the first three NMSSM SUGRA scenarios, the gamma-line
observable does not permit us to discriminate them. The distinct features of the
NMSSM Higgs sector are quite decoupled from the rest because $\lambda$ and
$\kappa$ are quite small for the first two scenarios. The situation is
different for the third scenario where $\lambda$ and $\kappa$ take
higher values and the NMSSM Higgs sector exhibits a different spectrum than the
MSSM one because, in this case, the singlet component mixes notably with the
doublets. Nevertheless, we recall that as the neutralino is a Majorana particle,
it forms a pseudoscalar state such that its couplings to scalar Higgses are 
suppressed at vanishing relative velocities, and therefore the LSP is mainly
sensitive to the pseudoscalar part of the Higgs spectrum. Moreover the
singlino-like neutralino is very heavy compared to the rest of the neutralino
spectra, such that it does not bring significant mixing. Thus
the couplings of the lightest neutralino does not depart much with respect to
the MSSM case. In addition a MSSM bino-like neutralino does not couple much to
Higgs at low $\tb$\footnote{The input value of $\tb$ is equal to 10 at the
$M_Z$ scale}, such that in the three bino scenarios considered the annihilations
are driven by box diagrams with sfermions, in particular staus. Let us now
comment the SUGRA-P4 scenario. The NMSSM benchmark point gives a LSP which is
almost a pure singlino and couples very feebly to other particles. Hence we
obtain a  very small rate, compared to the Higgsino-like LSP in the P4 MSSM
benchmark, since in this case the pair annihilation is quite efficient as the
LSP has $SU(2)$ quantum numbers.
\par\noi 
All in all, taking into account the latest limits provided by \texttt{FERMI}
\cite{fermigg}, none of these benchmarks points are excluded and far away from
\texttt{FERMI} sensitivity. As these benchmarks are representative of the NMSSM
phenomenology this means and confirms the \texttt{FERMI} collaboration statement
that the present limits on the gamma-ray line are too weak compared to the
typical cross sections for a conventional thermal NMSSM LSP. We also remark that
the situation is worse for a mostly singlino neutralino. This is quite generic
for a singlino $\neuto$ unless some specific mechanism is at play, we will come
back to this point later. As a final comment to this section we also observe
that the $\gamma Z$ mode gives generically smaller rates (in both the MSSM and
NMSSM) than the $\gamma\gamma$ one, as usually claimed, except for
the SUGRA-P4 MSSM scenario.
\subsection{GMSB}
We now turn to the GMSB benchmark points. These were taken once again from
\texttt{NMSSMTools\_2.3.5} package \cite{nmssmtools}. The most relevant ones
are depicted in Table~\ref{inputGMSB}.
\begin{table}[htbp]
 \begin{center}
  \begin{tabular}{|c|c|c|c|c|c|}
   \hline
   Parameter & GMSB-P1 & GMSB-P2 & GMSB-P3 & GMSB-P4 & GMSB-P5\\
   \hline
   $M_1$ & 473.32 & 473.42 & 136.69 & 136.91 &497.49 \\
   $M_2$ & 860.25 & 859.50 & 257.30 & 258.02 & 905.29\\
   $\mu$ & 1391.6 & 233.89 & 660.05 & 554.37 & 1363.0 \\
   $\tb$ & 8.4821 & 1.6277 & 1.5982 & 1.9000 & 50.055\\
   $ M_{\tilde l_{1,2}^R}$ & 692.31 & 689.42 & 138.22 & 133.34 & 622.57\\
   $ M_{\tilde l_3^R}$ & 686.00 & 689.20 & 138.17 & 133.28 & 429.74\\
   $ M_{\tilde u_{1,2}^R}$ & 231.81& 231.70 & 771.61& 777.47 & 2439.5\\
   $ M_{\tilde u_3^R}$ & 187.05 & 176.49 & 631.97 & 660.97 & 20361.8\\
   $ M_{\tilde d_{1,2}^R}$ & 228.40 & 228.27 & 767.45 & 773.63 & 2413.4\\
   $ M_{\tilde d_3^R}$ & 227.61& 228.23 & 767.36 & 773.52 & 2238.0\\
   $A_\mu$ & -424.24 & -321.01& -54.511 & -46.704 & -208.15\\
   $A_\tau$ & -423.01 & -318.15 & -54.333 & -46.573 & -184.48\\
   $A_b$ &-2115.1 & -1990.8 & -439.67 & -398.57 & -166.66\\
   $A_t$ &-1558.5 & -1265.4 & --328.32 & -314.23 & -143.41\\
   \hline
   \multicolumn{6}{|c|}{MSSM specific parameter} \\
  \hline
   $M_{A^0}$   & 1756.1 & 2884.8 & 875.97 & 714.730 & 1061.3\\
   \hline
   \multicolumn{6}{|c|}{NMSSM specific parameters} \\
   \hline
   $\lambda$ &0.002  & 0.5&0.6 & 0.6 &  0.01\\
   $\kappa$ & 0.0045 & 0.4351& 0.4540& 0.3969 & -0.0007\\
   $A_\lambda$ & -56.263 & -449.46& 23.012& 13.083&114.94\\
   $A_\kappa $ & -158.89 & -2278.7& 0.0315& 0.7354&0.0048\\
   \hline 
  \end{tabular}
\caption{{\em SUSY input parameters for the GMSB benchmarks. Masses are in
GeV.}\label{inputGMSB}}
 \end{center}
\end{table}\noi
The original GMSB phenomenology of the NMSSM has been studied in
\cite{ellwanger-gmsb}. The spectral lines results can be found in
Table~\ref{resGMSB}.
\begin{table}[htbp]
 \begin{center}
  \begin{tabular}{|c|c|c||c|}
 \hline 
 Model & $\mchi$ (GeV)& $\sigmav_{\gamma\gamma}\times
10^{30}[\mbox{cm}^3\, \mbox{s}^{-1}]$ & $\sigmav_{\gamma Z}\times
10^{30}[\mbox{cm}^3\, \mbox{s}^{-1}]$ \\
\hline
\multicolumn{4}{|c|}{GMSB-P1}  \\
\hline
NMSSM & 472.42 &$0.2915$ & $0.1176$\\
MSSM   & 472.42 & $0.2915$ & $0.1176$\\
\hline 
\multicolumn{4}{|c|}{GMSB-P2}  \\
\hline
NMSSM & 472.47 &$0.2981$ & $0.1185$\\
MSSM   & 472.47 & $0.2981$ & $0.1185$\\
\hline
\multicolumn{4}{|c|}{GMSB-P3}  \\
\hline
NMSSM & 133.01 &$12.862$& $6.0282$\\
MSSM   & 133.02 & $12.827$ & $6.0466$\\
\hline 
\multicolumn{4}{|c|}{GMSB-P4} \\
\hline
NMSSM & 132.50 &$14.490 $ & $7.1262$\\
MSSM   & 132.55 & $14.460$& $7.1544$\\
\hline
\multicolumn{4}{|c|}{GMSB-P5} \\
\hline
NMSSM & 196.84 &$2.6264\,10^{-10}$ & $5.5596\,10^{-11}$\\
MSSM   & 496.83 & $1.0884$ & $0.5055$\\
\hline
\end{tabular}
\caption{{\em Rates for the loop-induced annihilation of two
neutralinos into $\gamma\gamma/Z^0$ for the GMSB benchmarks.}\label{resGMSB}}
 \end{center}
\end{table}
\noi 
The GMSB-P1 and P2 benchmarks give both almost similar results with  $\mneuto
\simeq 472$ GeV. We observe no major discrepancy between the NMSSM and MSSM
results, even in scenario P2 where the NMSSM sector is not in the decoupling 
regime since $\lambda$ and $\kappa$ take quite high values. Nevertheless it
still has no impact because the LSP is 99\% bino-like and has MSSM-like
couplings. Note that for these two points the LSP mass is outside  the energy
range published in \cite{fermigg,vertongen11}. For benchmarks P3 and P4 the
differences between the NMSSM and the MSSM are tiny, already at the
level of the mass of the lightest neutralino $\mneuto$, in turn the rates are
slightly different. This is mostly due to the fact that the signal is computed
at a different center of mass energy. Even though it would correspond to
different signal/energy bins, we can anyway compare them since the mass
difference is quite small and require a high level of precision. Therefore
even discriminating between the two models is also extremely challenging, both
in $\gamma\gamma$ and $\gamma Z^0$ final states. As well as the mSUGRA benchmark
points where the LSP is mostly bino-like, the annihilation is driven by box
diagrams involving right-handed sfermions, and in particular the right-handed
sleptons since they possess the lower masses of the sfermion spectrum. The
benchmarks P3 and P4 give an enhanced rate because the right-handed slepton
masses are close to the mass of the neutralino. In addition the bino-like
neutralino in the P4 scenario has a substantial higgsino component, increasing
its couplings to the $Z^0$ boson. Finally the scenario P5 gives an extremely
tiny signal where the LSP is mostly singlino and the s-channel exchange of the
singlet pseudoscalar is extremely suppressed since its mass is around $1$ GeV.
\section{Spectral lines from light neutralino annihilation}
We have seen in the previous section that a mostly bino LSP in the NMSSM mimics
the MSSM result and gives quite low predictions with respect to the present
experimental limits. We propose in this section to investigate how the signal 
can be improved in the NMSSM, with features distinct from the MSSM.
Nevertheless, let us recall how we can increase the gamma-line signal with
features similar to the MSSM. In a generic neutralino case, one can look for a
point in parameter space where a $Z^0$ resonance is hit, with $\mneuto \simeq
M_Z/2 \simeq 45$ GeV, not forgetting the fact that the $\gamma Z^0$ channel
would require special care since we would produce the $Z^0$ boson at threshold.
Second a mostly wino, higgsino or a mixture of a bino with one of these states
would result in a signal with better prospects, like in the MSSM. However such
type of neutralinos must be quite heavy, especially for wino- or Higgsino-like
neutralinos. Indeed, if one wants to reproduce the correct CDM abundance,
typically the masses must be of order the TeV scale\footnote{Scenarios with
heavy MSSM wino-like or Higgsino-like LSP also require special care for the
gamma-line computation, in particular the treatment of the Sommerfeld
singularity, see the pioneer work of \cite{Nojiri-gammaray-coulomb}}. 
Motivated by results indicative of light dark matter, we focused on the case
of a NMSSM light neutralino, which can have very different properties than the
one of the MSSM. It is very difficult in the MSSM to have a very light
neutralino which do not overclose the Universe and at the same time fulfills the
collider constraints \cite{daniel-nmssm,daniel-mssm,fornengo11,calibbi11}.
Reconciliating these two criteria can be provided by the NMSSM more easily.
Indeed the lightest of the two CP-odd Higgs bosons can be much lighter than the
single CP-odd Higgs boson of the MSSM without violating collider constraints.
Another feature that cannot be reproduced in the MSSM is obviously the one where
the LSP is mostly singlino. We have seen that on general grounds the signal is
very low for such a $\neuto$. Nevertheless, just like with a $Z^0$ boson we can
enhance the signal by hitting a resonance, this time with one pseudoscalar
Higgs. This mechanism has already been pointed out in \cite{profumogg}, and is
also invoked to give the right relic density when the LSP possesses an important
singlino component \cite{daniel-nmssm,belanger-nmssm}. 
We therefore investigated to which extent the signal can be enhanced for such a
mechanism with neutralinos lighter than 15 GeV, and if the present limits can be
used to further constrain the parameter space found in \cite{daniel-nmssm}.
 Such light neutralinos would evade the present
constraints from the published \texttt{FERMI} data on gamma-lines since their
threshold is around 30 GeV \cite{fermigg}. However Vertongen
and Weniger \cite{vertongen11} have derived a broader energy range which
starts from 1 GeV and extends up to 300 GeV, see Table.~3 of \cite{vertongen11}.
As a consequence we will only discuss how to increase the $\gamma\gamma$ signal
since the $\gamma Z^0$ channel is closed in this range, and we will compare our
predicted rates with the limits given in \cite{vertongen11}. 
\subsection{Light singlet pseudoscalar Higgs resonance}
\begin{figure}[htbp]
\begin{center}
 \epsfig{file=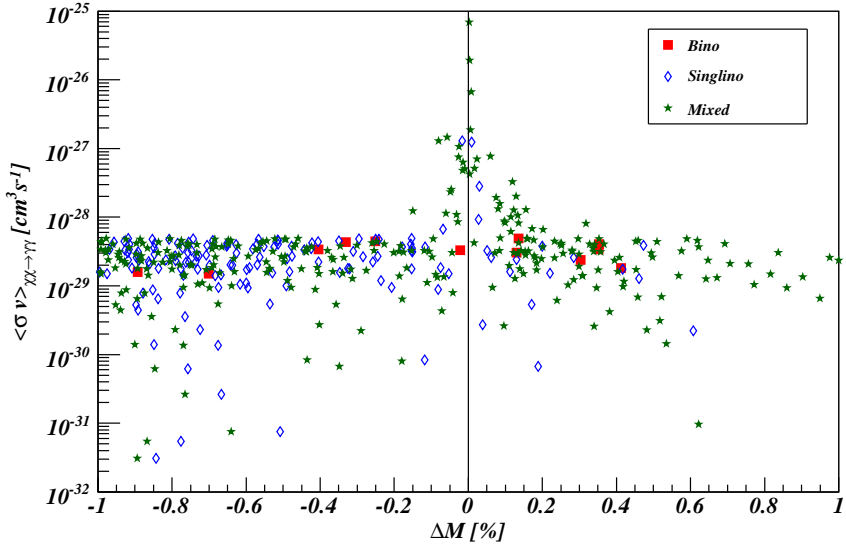,clip=,width=0.7\textwidth}
\caption{\label{resA} \em Gamma-ray line rate with respect to the mass
difference $\Delta M = (2\mneuto - m_{A_1})/m_{A_1}$ for several composition of
the neutralino (see text for details).}
\end{center}
\end{figure}

\begin{figure}[htbp]
 \begin{center}
  \epsfig{file=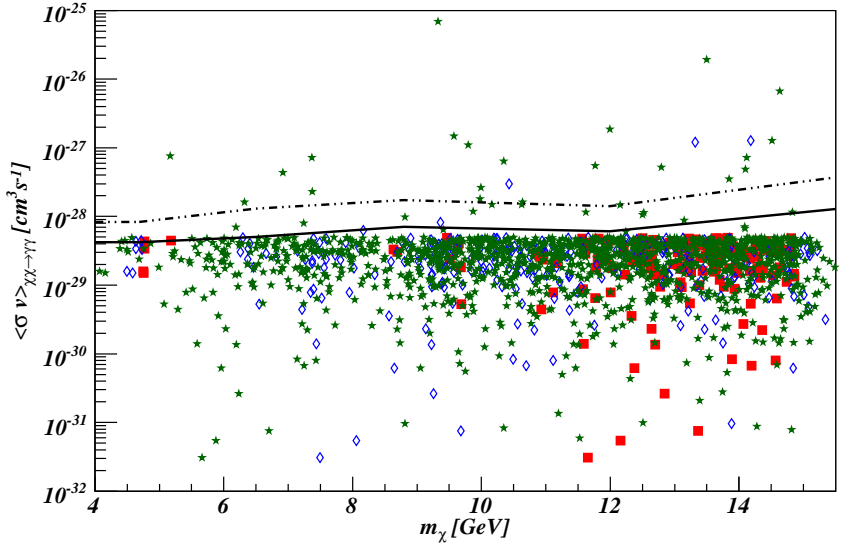,clip=,width=0.7\textwidth}
\caption{\label{resA-excl}\em Gamma-ray line rate with respect to $\mchi$.
The limits of \cite{vertongen11} are drawn for the halo (dashed
dotted line) and the galactic center (solid line). The conventions for the
neutralino composition follow those of Fig.~\ref{resA}.}
 \end{center}
\end{figure}

Light neutralinos in the NMSSM have been studied for example in
\cite{daniel-nmssm,gunion-dmlight, guniondm-10}. A singlino-like $\neuto$
emerges when $\kappa \ll 1$ and to ensure at the same time a light singlet
pseudoscalar the parameter $A_\kappa$ should be small either (in the decoupling
limit $m_{A_1}^2 \simeq - 3 \kappa A_\kappa s$). Light singlino-like LSP are
found mainly in two regions of parameter space : one at small $\lambda$ with
very small $|\kappa|$, and another at large $\lambda$ with slightly larger
$|\kappa|$ allowed, see \cite{gunion-dmlight}. To hit a light singlet
pseudoscalar Higgs resonance we obviously need $\mneuto \simeq m_{A_1}/2$.
This can be achieved with a pure singlino-like neutralino but, in practice,
neutralinos fulfilling this criteria and accounting for the present cosmic
abundance of dark matter measured by \texttt{WMAP} are mixed states,
like a bino-singlino or bino-higgsino like neutralino. In any case the
neutralino obviously needs some substantial component related to the Higgs
sector
to couple to the pseudoscalars. Such light pseudoscalars and neutralinos can be
motivated in models were an approximate $U(1)$ R-symmetry or $U(1)$ Peccei-Quinn
symmetry holds (see \cite{gunion-dmlight} and references therein) and therefore
their ''lightness`` is more ''natural`` than in the MSSM. To see in which extent
the signal can be boosted by this mechanism we performed a scan over the
parameter spaced spanned by the 11 parameters 
\begin{equation*}
 M_1, M_2, \mu, \tb, \lambda,\kappa,A_\lambda,A_\kappa, A_t, m_{\tilde
l}, m_{\tilde q}
\end{equation*}\noi 
as in \cite{daniel-nmssm} and computed the rate
$\sigmav_{\chi\chi\ra\gamma\gamma}$ as only this channel is open. The parameters
$m_{\tilde l}$ and $m_{\tilde q}$ are common masses to the sleptons and squarks
respectively and are taken in the same range as in \cite{daniel-nmssm}. However
to find these points we did not apply any constraints, we just tried  to find
points in parameter space which where minimizing the difference
\begin{equation}
 \Delta M = \frac{2\mneuto - m_{A_1}}{m_{A_1}}
\end{equation}\noi 
such that we observed an enhanced signal, and with $\mneuto \leq 15.5$ GeV. No
width to the pseudoscalar $A_1$ propagator was introduced during the
calculation, since we did not encounter any numerical problems. This is due to
the fact that the width of the pseudoscalar is extremely narrow (numerically
$10^{-7} \lesssim \Gamma_{A_1} \lesssim 10^{-4}$). Anyway, as a check, we added
a width to the propagator of the lightest pseudoscalar for the points giving the
highest rates, i.e, very close to the resonance, and we observed a maximum
decrease of $\sim$ 20\% of the signal, so these points would still be excluded.
As expected, the effect of the width was
less and less important (reaching the percent level) the more we were going away
from the resonance, where more points are allowed. The effects of the width for
these allowed points is therefore lower than the present experimental
accuracy. The result of this scan is displayed in Fig.~\ref{resA}.
\noi 
Red squared points are scenarios where the LSP is more than 90\% bino, blue
diamond-shaped points more than 90\% singlino and the green stars are a mixture
of both. We displayed only the region where $| \Delta M| \leq 1\%$.
We can see that for this mechanism to boost significantly
the signal a high degree of fine-tuning of $\Delta M$ is required. Very few
bino-like neutralinos scenarios were found and give a signal lower than
$10^{-28}\,\mbox{cm}^3\,\mbox{s}^{-1}$, showing that this mechanism is specific
to the NMSSM. Most of the points possess also a significant Higgsino component
which ranges from the percent level to approximately 15\%. Notice also
that as soon as $|\Delta M | \gtrsim 0.2\%$ the rate significantly decreases
and more points were produced with $m_{A_1} \geq 2 \mneuto$. We
observe that in the most extreme case we can reach a signal as large as
$10^{-25}\,\mbox{cm}^3\,\mbox{s}^{-1}$, for a mixed $\neuto$, which is 88\%
singlino and 9\% Higgsino while $\Delta M \simeq 0.001 \%$. However recall that
we did not apply any constraining criteria for this scan. In
Fig.~\ref{resA-excl} we display the result of this scan with respect to the
limits given in \cite{vertongen11}. This time we picked up points with $|\Delta
M| \leq 15\%$, this selected the mass window $ 4 \leq \mneuto \leq
15.5~\mbox{GeV}$.
\noi
 We see that the highly fine-tuned points (with $|\Delta M| \leq 0.2\%$ and
$\sigmav_{\gamma\gamma} \gtrsim 10^{-28}~\mbox{cm}^3/s$) would be
excluded by the spectral lines searches. However we found that most of the
points are just below the \texttt{FERMI} sensitivity, especially concerning the
galactic center observation. Therefore an increase by at least one order of
magnitude would much more constrain the s-channel pseudoscalar resonance, if no
signal is reported. It is also worth to note that the highly fine-tuned
scenarios tend also to give a very low relic density\footnote{Nevertheless we
found some points satisfying this criteria.}, hence giving up the possibility of
the LSP to account for the actual amount of dark matter. Moreover, as we will
discuss in the next section, the gamma-rays from dwarf spheroidal
galaxies (dSph)\cite{daniel-astrolim,kaplinghat06} give more powerful
constraints. 
\subsection{Gamma-ray lines as a further constraint on the NMSSM parameter
space.}
Interestingly the studies led by \cite{daniel-nmssm,cumberbatch11} found that
light NMSSM neutralinos that are compatible with \texttt{CoGeNT} and satisfy all
the constraints they applied (colliders, relic density of DM, flavour
observables) are accompanied by a
light pseudoscalar and/or scalar singlet. Generally the aforementioned studies
showed that the highest elastic scattering cross-sections are reached for low
values of $m_{H_1}$ and can satisfy the WMAP constraint when $m_{H_1} \simeq 2
\mneuto$, because in this case the pair annihilation of neutralinos is enhanced
by a s-channel resonance. Obviously the gamma-ray lines cannot constrain
such scenarios as this channel is suppressed for low-velocities. Scenarios that
could be constrained are therefore rather those where $\mneuto \simeq
m_{A_1}/2$. 
\par\noi 
The authors of \cite{daniel-astrolim} computed the secondary gamma rays produced
in dSph from the pair annihilation of DM particles into quarks and/or taus
which subsequently hadronise and decays into pions, finally decaying into
photons. The authors then compared their findings with the \texttt{FERMI} 95 \%
limits on gamma-ray emission from dSph \cite{fermi-dsph}. The authors provided
us with 14 points of their MCMC scan giving a large pair annihilation cross
section but safe with respect to dSph limits and direct detection searches.
The points are sampled in each bins of $\mneuto$ between 1 and 15 GeV. We then
used these input parameters to evaluate the rate
of the gamma-lines to see if the corresponding limits derived in
\cite{vertongen11} from the \texttt{FERMI} data on spectral lines \cite{fermigg}
could further constrain these scenarios. The mass spectrum of these parameter
points fall into the range where $m_{A_1}$ is not far from $2\mneuto$. However
models for which the mass difference between the LSP and the pseudoscalar is
highly fine-tuned are already excluded by the Draco limits
\cite{daniel-astrolim}. 
\begin{table}[h]
 \begin{center}
  \begin{tabular}{|c|c|c|}
   \hline
   $\mneuto $ & $\sigmav_{\chi\chibar\ra q\bar q, \tau\bar\tau}
\times 10^{27}$ & $\sigmav_{\gamma\gamma}\times 10^{30}$ \\
  $[\mbox{GeV}]$ & $[\mbox{cm}^3\, \mbox{s}^{-1}]$ & $[\mbox{cm}^3\,
\mbox{s}^{-1}]$ \\
   \hline 
  \hline
  0.976 & 	0.209 & 0.00008\\
  \hline 
  2.409 & 0.297 &  0.00267\\
  \hline 
  3.342 & 0.345 & 0.00345\\
  \hline 
  4.885 & 3.298 & 0.00262\\
  \hline
  5.626 & 5.389 & 0.00410\\
  \hline 
  6.551 & 3.547 & 0.00427\\
  \hline
  7.101 & 2.425 & 0.00664\\
  \hline 
  8.513 & 2.161 & 0.00220\\
  \hline 
  9.274 & 2.497 & 0.00655\\
  \hline 
  10.27 & 2.323 & 0.01881\\
  \hline 
  11.50 & 2.575 & 0.02456\\
  \hline
  12.74 & 3.224 & 0.02003\\
  \hline 
  13.51 & 9.571 & 0.17487\\
  \hline 
  14.48 & 148.4 & 2.87500\\
  \hline
  \hline
  \end{tabular}
\caption{\em Comparison between dark matter annihilation into quarks
and/or taus (values are taken from \cite{daniel-nmssm}\label{tabDSph}) and the
loop-induced one into photons in the NMSSM for each of the bins
between 1 and 14 GeV.}
 \end{center}
\end{table}\par\noi 
We therefore do not expect very bright signals. We collected in
Table~\ref{tabDSph} the total annihilation cross section of neutralinos into
quarks and/or taus, taken from \cite{daniel-astrolim}, with our results
concerning the annihilation into two photons, since here again the range of
$\mneuto$ studied is too low for the production of the $\gamma Z^0$ final state.
As can be seen from Tab.~\ref{tabDSph}, the gamma-line rate is at most four
orders of magnitude lower than the rate coming from other indirect search
channels. Of course since the direct annihilation into photons is
loop-suppressed we cannot expect that it can significantly contribute to the
total annihilation cross section producing gamma rays. For the computation  of
gamma ray fluxes emerging from dark matter annihilation in dwarf spheroidal
galaxies, the ''direct`` annihilation into photons can then safely be neglected.
If we now compare the results in Tab.~\ref{tabDSph} with the limits of
\cite{vertongen11}, which are at best of order $10^{-29}\,\mbox{cm}^3/ \mbox{s}$
for a very light neutralino (less than approximately 5 GeV), none of these
points is further constrained. Most of the models exhibits a spectrum with a
mass difference $\Delta M$ too high to enhance significantly the signal, and
relatively low mass sleptons. The point with $\mneuto
= 14.48$ GeV gives the best signal because $\Delta M \simeq 0.5 \%$ with
a LSP at 92\% singlino and 8\% Higgsino. A sensitivity increased by at
least two order of magnitude would be needed to exclude this point with respect
to the constraints from gamma-lines searches. In consequence we conclude  that
the present limits on gamma ray lines can constrain models with a very
low mass difference $\Delta M$ but they are not competing with the gamma ray
searches from dwarf spheroidal galaxies anyway. This can be easily understood by
the fact that photons produced from processes like $\neuto\neuto \ra q \bar q /
\tau \bar\tau$ are leading-order dominated and experiments have not yet
reached a sensitivity down to loop effects, like the spectral lines.

\section*{Conclusion}
The mono-energetic gamma ray line signal has spectacular features : a clear
''smoking-gun`` signature and a direct relation to the mass of the dark matter
particle. Moreover it do not suffer from astrophysical uncertainties and
depend only on the assumption of the dark matter halo. However discriminating
it from the overwhelming astrophysical background (supernov\ae{}, pulsars,
cosmic-rays...) is experimentally extremely challenging and requires a fine
energy resolution. Furthermore the inclusive channels $\chi \chi \ra q \bar q
(WW,ZZ) $ produce a featureless continuous spectrum of gammas coming from the
decay or hadronisation of the final state, only cut-off at a maximum energy
corresponding to the mass of the DM particle. This contribution is stronger than
the monochromatic photons, since it is a tree-level dominated process. In this
paper we evaluated numerically in the NMSSM framework the process $\neuto\neuto
\ra \gamma \gamma$ and provided for the first time result for $\neuto \neuto \ra
\gamma Z^0$. This is the first implementation of the NMSSM framework in
\SloopS~and paves the way for a future renormalisation of each sector, like what
was done for the MSSM \cite{baro08,baro09}. We performed the calculation at
$v=0$ and checked the result with respect to gauge invariance thanks to an NMSSM
extension of the non-linear gauge-fixing used in \cite{baro08}. As a further
check we compared the NMSSM result in the ''MSSM-limit`` and the pure MSSM one
and found an excellent agreement. As a first application we computed the
$\gamma\gamma(Z^0)$ signal for several benchmarks found in the literature
\cite{bmp-nmssm,ellwanger-gmsb}. We compared with an ''equivalent'' MSSM
spectrum and the limits given by \texttt{FERMI} \cite{fermigg} and in
\cite{vertongen11}. We found that unless the lightest neutralino has a
significant singlino component, the two models produce very similar rates and
discriminating between them requires a high level of experimental precision.
This is particular true if in both cases the LSP is bino-like. However, if the
neutralino is to be the dark matter candidate and light, discriminating the
NMSSM from the MSSM should be easier since light neutralinos and Higgs scalars
are more ``natural'' in the NMSSM, thanks to an approximate Peccei-Quinn
symmetry. A low mass WIMP is therefore easier to accommodate in the NMSSM than
the MSSM. In the NMSSM the Higgs sector is more peculiar and its related
collider observables are expected to give also more spectacular signatures,
hence are more prone to unveil the singlet component of the Higgs potential
\cite{nmssmrev-ellwanger}. Therefore dark matter related observables in the
NMSSM are also distinct when the singlet component is at play both in the
neutralino and Higgs sector. Furthermore, as far as the annihilation
$\neuto\neuto \ra \gamma\gamma(Z^0)$ is concerned, this observable can only be
sensitive to the pseudoscalar Higgs states, since at $v=0$ scalar couplings are
suppressed. In this particular case, such regions of parameter space are only
reachable in the NMSSM. A noticeable example of such a mechanism is
self-annihilation of dark matter particles through a s-channel resonance with a
light pseudoscalar. We investigated such a mechanism with a light $\neuto$ for
the $\gamma\gamma$ final state and observed that the signal can be very bright
when extremely close to the resonance. A singlino-like LSP not fulfilling this
feature would on the contrary give a very low rate. These kind of highly
fined-tuned scenarios would in fact be excluded taking into account the limits
derived by the authors of \cite{vertongen11}. However, if this mechanism is
actually at play for the indirect detection of primaries photons, it is also the
case for both the diffuse photon spectrum and relic density calculations.
Indeed, if ones wants to match the CDM abundance, this fixes the pair
annihilation rate to be of order the canonical value of $\sigmav \sim 3\times
10^{-27}~\mbox{cm}^3/\mbox{s}$. Interestingly, this value is in the
ballpark of the \texttt{FERMI} sensitivity. Unfortunately, as the gamma
lines is a loop suppressed process, the interesting sensitivity is one or
two order of magnitude less, in the most favourable scenarios, like the
pseudoscalar resonance. Outside these
regions, limits on the monochromatic gamma lines are not very constraining, in
particular, on the astrophysical level, with respect to the ones from dwarf
spheroidal galaxies. Furthermore, even concerning the pseudoscalar resonance,
the dSph put also some more stringent constraints on this mechanism, since
it is also at play there. The sensitivity of experiments concerning the spectral
lines still need to be improved by several orders of magnitude to be compelling.
The \texttt{FERMI} collaboration claims that their limits on the spectral lines
will be significantly improved by the time the mission is finished and we can
expect a possible discovery or, in case of a null result, that more featureless
regions of the NMSSM could be excluded.
\section*{Acknowledgements}
The authors would like to thank G.~Bélanger and F.~Boudjema for bringing our
attention to this topic and very instructive comments. We would also like to 
thank U.~Nierste and P.~Anghel-Vasilescu for proofreading the
manuscript. GC thanks F.~Domingo for helpful conversations,
D.~A.~Vazquez for useful discussion and for providing us with points presented
in \cite{daniel-nmssm}. The Figures were created with the help
of \texttt{JaxoDraw} \cite{jaxodraw}. This work was supported in part by the
GDRI-ACPP of CNRS. The work of GC is supported by BMBF grant
05H09VKF. 

\renewcommand{\theequation}{A.\arabic{equation}} 
  \setcounter{equation}{0}  
\section*{Appendix : NMSSM BRST transformations for the scalar fields}
To generate the ghost Lagrangian through the BRST transformations starting from
the non-linear gauge-fixing in Eq.~(\ref{gaugefixing}) we need, in addition to
the BRST transformations for the gauge fields which can be found in
\cite{grace-1loop}, the BRST transformations of the scalar fields from the
gauge transformation of the Higgs doublets.
 We can parametrise the fields of the Higgs sector in the following way
\begin{equation}
 H_d = \begin{pmatrix}
         (v_d + \phi_d + i \varphi_d)/\sqrt{2} \\ -\phi_d^-
        \end{pmatrix},~
 H_u = \begin{pmatrix}
         \phi_u^+ \\ (v_u + \phi_u + i \varphi_u )/\sqrt{2}
        \end{pmatrix},~
 S =  (v_s + \phi_s + i \varphi_s )/\sqrt{2}
\end{equation}\noi 
The following unitary rotations matrices turn the gauge eigenstates to the
physical ones
\begin{eqnarray}
 \left(\begin{array}{c} \phi_{1}^{\pm}\\ \phi_{2}^{\pm}\end{array}\right)
&=&\left(\begin{array}{cc} c_{\beta} & -s_{\beta}\\
s_{\beta} & c_{\beta}\end{array}\right)\left(\begin{array}{c}
G^{\pm}\\H^{\pm}\end{array}\right)  \, , \\
\left(\begin{array}{c} \phi_1 \\ \phi_2 \\ \phi_s\end{array}\right)
&=&\left(\begin{array}{ccc}
 Z^h_{11} & Z^h_{12} & Z^h_{13}\\
 Z^h_{21} & Z^h_{22} & Z^h_{23}\\
 Z^h_{31} & Z^h_{32} & Z^h_{33}\\
\end{array}\right) \left(\begin{array}{c} H_1 \\
H_2 \\ H_3\end{array}\right) \, , \\
\label{rotpseudoscal}
\left(\begin{array}{c} \varphi_1 \\ \varphi_2 \\ \varphi_s\end{array}\right)
&=&\left(\begin{array}{ccc}
 Z^a_{11} s_\beta  & Z^a_{21} s_\beta &  c_\beta \\
 Z^a_{11} c_\beta & Z^a_{21} c_\beta  & -s_\beta\\
Z^a_{12}& Z^a_{22} & 0 \\
\end{array}\right) \left(\begin{array}{c} A_1 \\
A_2 \\ G^0\end{array}\right) \, , \\
\end{eqnarray}\noi 
where we have first rotated with an angle $\beta$ in the pseudoscalar sector to
single out the goldstone mode $G^0$ from the physical CP-odd states in
Eq.~(\ref{rotpseudoscal}) \cite{nmssmrev-ellwanger}. As the BRST transformations
are closely linked to gauge transformations we have for the singlet components,
\begin{equation}
\delta_{\mbox{\tiny BRS}} \phi_s = \delta_{\mbox{\tiny BRS}} \varphi_s = 0
\end{equation}\noi 
Provided the rotations matrices we have then a relation between the BRST
transformations of the physical eigenstates once the ones for the gauge
eigenstates are known,
\begin{eqnarray}
 \left(\begin{array}{c} \delta_{\mbox{\tiny BRS}}
G^{\pm}\\\delta_{\mbox{\tiny BRS}} H^{\pm}\end{array}\right)
&=&\left(\begin{array}{cc} c_{\beta} & s_{\beta}\\
-s_{\beta} & c_{\beta}\end{array}\right) \left(\begin{array}{c}
\delta_{\mbox{\tiny BRS}}\phi_{1}^{\pm}\,\\
\delta_{\mbox{\tiny BRS}} \phi_{2}^{\pm}\end{array}\right) \\
 \left(\begin{array}{c} \delta_{\mbox{\tiny BRS}}H_1 \\ \delta_{\mbox{\tiny
BRS}}H_2 \\ \delta_{\mbox{\tiny BRS}}H_3\end{array}\right)
&=&\left(\begin{array}{ccc}
 Z^h_{11} & Z^h_{21} & Z^h_{31}\\
 Z^h_{12} & Z^h_{22} & Z^h_{32}\\
 Z^h_{13} & Z^h_{23} & Z^h_{33}\\
\end{array}\right)\left(\begin{array}{c} \delta_{\mbox{\tiny BRS}}\phi_1
\\\delta_{\mbox{\tiny BRS}} \phi_2 \\
0\end{array}\right) \, \\
\left(\begin{array}{c} \delta_{\mbox{\tiny BRS}} A_1 \\ \delta_{\mbox{\tiny
BRS}}A_2 \\ \delta_{\mbox{\tiny BRS}}G^0\end{array}\right)
&=&\left(\begin{array}{ccc}
 Z^a_{11} s_\beta & Z^a_{21} s_\beta & Z^a_{12} \\
Z^a_{21} c_\beta &Z^a_{21} c_\beta & Z^a_{22} \\
 c_\beta & -s_\beta & 0 \\
\end{array}\right)\left(\begin{array}{c} \delta_{\mbox{\tiny BRS}}\varphi_1 \\
\delta_{\mbox{\tiny BRS}}\varphi_2 \\
0\end{array}\right) \,
\end{eqnarray}\noi 
With
\begin{eqnarray}
 \delta_{\mbox{\tiny BRS}} \phi_1^\pm & = & \mp \frac{ig}{2}c^\pm [ v_1 +
(Z^h_{11} H_1 +
Z^h_{12} H_2+ Z^h_{13} H_3) \mp i (\sinb(Z^a_{11} A_1 + Z^a_{21} A_2) +
\cosb G^0) ]\non
\\
&  &\mp i e \left(c^A - \frac{s_w^2 - c_w^2}{2 s_w c_w} \right)[c_\beta G^\pm -
s_\beta H^\pm]\\
\delta_{\mbox{\tiny BRS}} \phi_2^\pm & = &\mp \frac{ig}{2}c^\pm [ v_2 +
(Z^h_{21} H_1 +
Z^h_{22} H_2+ Z^h_{23} H_3) \pm i (\cosb(Z^a_{11} A_1 + Z^a_{21} A_2) - \sinb
G^0) ]\non
\\
&  &\mp ie \left(c^A - \frac{s_w^2 - c_w^2}{2 s_w c_w} \right)[s_\beta G^\pm +
c_\beta H^\pm] \\
\delta_{\mbox{\tiny BRS}} \phi_1^0 & = & + \frac{ig}{2}[c_\beta (G^- c^+ -  G^+
c^- ) - s_\beta(H^- c^+ - H^+ c^-)] \non \\
&  & +\frac{e}{2c_w s_w} c^Z [\sinb(Z^a_{11} A_1 + Z^a_{21} A_2) + \cosb G^0] \\
\delta_{\mbox{\tiny BRS}} \phi_2^0 & = & + \frac{ig}{2}[s_\beta(G^- c^+ -  G^+
c^- ) + c_\beta (H^- c^+ - H^+ c^-)] \non \\
& & - \frac{e}{2c_w s_w} c^Z [\cosb(Z^a_{11} A_1 + Z^a_{21} A_2) - \sinb G^0] \\
\delta_{\mbox{\tiny BRS}} \varphi_1^0 & = & + \frac{g}{2}[c_\beta (G^- c^+ + 
G^+ c^- ) - s_\beta (H^- c^+ + H^+ c^-)] \non \\
& & - \frac{e}{2c_w s_w} c^Z [ v_1 + Z^h_{11} H_1 + Z^h_{12} H_2 + Z^h_{13}
H_3]\\
\delta_{\mbox{\tiny BRS}} \varphi_2^0 & = & - \frac{g}{2}[s_\beta (G^- c^+ + 
G^+ c^- ) + c_\beta (H^- c^+ + H^+ c^-)] \non \\
& & + \frac{e}{2c_w s_w} c^Z [ v_2+Z^h_{21} H_1 + Z^h_{22} H_2 + Z^h_{23} H_3]
\end{eqnarray}



\clearpage
\newpage
\begin{thebibliography}{10}

\bibitem{cdmsII10}
[\texttt{CDMS-II} Collaboration], Z.~Ahmed {\it et al.}, {\it Science} {\bf327}
  (2010) 1619, arXiv:0912.3592 [astro-ph].

\bibitem{cogent10}
[\texttt{CoGeNT} Collaboration], C.~E.~Aalseth {\it et al.}, {\it Phys. Rev.
  Lett.} {\bf 106} (2011) 131301, arXiv:1002.4703 [astro-ph.CO].

\bibitem{cogent11}
[\texttt{CoGeNT} Collaboration], C.~E.~Aalseth {\it et al.}, {\it
Phys. Rev. Lett.} {\bf 107} (2011) 141301, arXiv:1106.0650
  [astro-ph.CO].

\bibitem{cresst11}
[\texttt{CRESST} Collaboration], G.~Angloher {\it et al.}, arXiv:1109.0702
  [astro-ph.CO].

\bibitem{xenon100-11}
[\texttt{XENON100} Collaboration], E.~Aprile {\it et al.}, arXiv:1104.2549
  [astro-ph.CO].

\bibitem{dama10}
R.~Bernabei, P.~Belli, F.~Capella {\it et al.}, {\it Eur. Phys. J.} {\bf C67}
  (2010) 39, arXiv:1002.1028 [astro-ph.GA].

\bibitem{daniel-nmssm}
D.~A.~Vazquez, G.~Belanger, C.~Boehm, A.~Pukhov and J.~Silk, {\it Phys. Rev.}
  {\bf D82} (2010) 115027, arXiv:1009.4380 [hep-ph].

\bibitem{daniel-mssm}
D.~A.~Vazquez, G.~Belanger, C.~Boehm, arXiv:1108.1338 [hep-ph].

\bibitem{das10}
D.~Das, U.~Ellwanger, {\it JHEP} {\bf 1009} (2010) 085, arXiv:1007.1151
  [hep-ph].

\bibitem{draper10}
P.~Draper, T.~Liu, C.~E.~M.~Wagner, L.~Wang, H.~Zhang, {\it Phys.~Rev.~Lett.}
  {\bf 106} (2011) 121805, arXiv:1009.3963 [hep-ph].

\bibitem{kappl10}
R.~Kappl, M.~Ratz, M.~W.~Winkler, {\it Phys.~Lett.} {\bf B695} (2011) 169,
  arXiv:1010.0553 [hep-ph].

\bibitem{cao11}
J.~Cao, K.~Hikasa, W.~Wang, J.~M.~Yang, {\it Phys.~Lett.} {\bf B703}, (2011)
  292, arXiv:1104.1754 [hep-ph].

\bibitem{fornengo11}
N.~Fornengo, S.~Scopel, A.~Bottino, {\it Phys.~Rev.} {\bf D83} (2011) 015001,
  arXiv:1011.4743 [hep-ph].

\bibitem{calibbi11}
L.~Calibbi, T.~Ota, Y.~Takanishi, {\it JHEP} {\bf 1107} (2011) 013,
  arXiv:1104.1134 [hep-ph].

\bibitem{nmssmrev-ellwanger}
U.~Ellwanger, C.~Hugonie, A.~M.~Teixeira, {\it Phys. Rept.} {\bf 496} (2010) 1,
  arXiv:0910.1785 [hep-ph].

\bibitem{profumogg}
F.~Ferrer, L.~Krauss, S.~Profumo, {\it Phys.\ Rev. } {\bf D74} (2006), 115007,
  hep-ph/0609257.

\bibitem{fermigg}
A. Abdo et al., {\it Phys. Rev. Lett.} {\bf 104} (2010) 091302, arXiv:1001.4836
  [astro-ph.HE].

\bibitem{ams}
[\texttt{AMS} Collaboration], S.~P.~Ahlen {\it et.al}, {\it
  Nucl.~Instrum.~Meth} {\bf A350} (1994) 351\\ J.~Alcaraz, {\it et.al}, {\it
  Nucl.~Instrum.~Meth} {\bf A478}, (2002) 119 \\ \url{http://ams.cern.ch/}.

\bibitem{lsptogg}
L.~Bergstr\"om, P.~Ullio, \textit{Nucl. Phys.} {\bf 504} (1997) 27,
  hep-ph/9706232;\\ Z.~Bern, P.~Gondolo, M.~Perelstein, \textit{Phys. Lett.}
  {\bf B411} (1997) 86, hep-ph9706538.

\bibitem{lsptozg}
P.~Ullio, L.~Bergstrom, {\it Phys.~Rev.} {\bf D57} (1998) 1962; hep-ph/9707333.

\bibitem{lsptoggfernand}
G.J.~Gounaris, J.~Layssac, P.I. Porfyriadis, F.M.~Renard, {\it Phys.~Rev.} {\bf
  D69} (2004) 075007; hep-ph/0309032.

\bibitem{boudjema05}
F.~Boudjema, A.~Semenov, D.~Temes, \textit{Phys. Rev.} {\bf D72} (2005) 055024,
  hep-ph/0507127.

\bibitem{bmp-nmssm}
A.~Djouadi {\it et.~al.}, {\it JHEP} {\bf 0807} (2008) 002, arXiv:0801.4321
  [hep-ph].

\bibitem{ellwanger-gmsb}
U.~Ellwanger, C.-C.~Jean-Louis, A.~M.~Teixeira, {\it JHEP} {\bf 0805} (2008)
  044, arXiv:0803.2962 [hep-ph].

\bibitem{baro08}
N.~Baro, F.~Boudjema, A.~Semenov, \textit{Phys. Rev.} \textbf{D78} (2008)
  115003, arXiv:0807.4668 [hep-ph].

\bibitem{baro09}
N.~Baro, F.~Boudjema, \textit{Phys. Rev.} {\bf D80} (2009) 076010,
  arXiv:0906.1665[hep-ph].

\bibitem{baro07}
N.~Baro, F.~Boudjema, A.~Semenov, \textit{Phys. Lett.} {\bf B660} (2008) 550,
  arXiv:0710.1821 [hep-ph].

\bibitem{barosusy09}
N.~Baro, G.~Chalons, Sun~Hao, Proceedings of SUSY09, arXiv:0909.3263 [hep-ph].

\bibitem{chalons09}
N.~Baro, F.~Boudjema, G.~Chalons, Sun Hao, \textit{Phys. Rev.} \textbf{D81}
  015005 (2010), arXiv:0910.3293 [hep-ph].

\bibitem{belanger-nmssm}
G.~Belanger, F.~Boudjema, C.~Hugonie, A.~Pukhov, A. Semenov, {\it JCAP} {\bf
  0509} (2005) 001, hep-ph/0505142.

\bibitem{Hao09}
F.~Boudjema, N.~ Le Duc, H. Sun Hao, M.~Weber, {\it Phys. Rev.} {\bf D81}
  (2010) 073007, arXiv : 0912.4234 [hep-ph].

\bibitem{lanhep}
A.~Semenov, hep-ph/9608488; \\ A.~Semenov, \textit{Nucl. Inst. Meth. and Inst.}
  {\bf A293} (1997) 293;\\ A.~Semenov, \textit{Comp. Phys. Commun.} {\bf 115}
  (1998) 124;\\ A.~Semenov, hep-ph/0208011; \\ A.~Semenov, {\em Comput. Phys.
  Commun.} {\bf 180} (2009) 431, arXiv:0805.0555 [hep-ph].

\bibitem{formcalc}
T.~Hahn, M.~Perez-Victoria, \textit{Comp. Phys. Commun.} {\bf 118} (1999) 153,
  hep-ph/9807565;\\ T.~Hahn, hep-ph/0406288; hep-ph/0506201.

\bibitem{chopin-nlg}
F.~Boudjema, E.~Chopin, \textit{Z. Phys.} {\bf C73} (1996) 85, hep-ph/9507396.

\bibitem{grace-1loop}
G.~B\'{e}langer, F.~Boudjema, J.~Fujimoto, T.~Ishikawa, T.~Kaneko, K.~Kato,
  Y.~Shimizu, \textit{Phys. Rep.} {\bf 430} (2006) 117, hep-ph/0308080.

\bibitem{looptools}
T. Hahn, {\tt LoopTools}, \url{http://www.feynarts.de/looptools/}.

\bibitem{bergstrom94gg}
L.~Bergstrom, J.~Kaplan, {\it Astropart. Phys.} {\bf 2} (1994) 2,
  hep-ph/9403239.

\bibitem{nmssmtools}
U.~Ellwanger, J.~F.~Gunion, C.~Hugonie, {\it JHEP} {\bf 0502} (2005) 066,
  arXiv:hep-ph/0406215\\ U.~Ellwanger, C.~Hugonie, {\it Comput.~Phys.~Commun.}
  {\bf 175} (2006) 290, arXiv:hep-ph/0508022\\ \url{
  http://www.th.u-psud.fr/NMHDECAY/nmssmtools.html}.

\bibitem{vertongen11}
G.~Vertongen, C.~Weniger, {\it JCAP} {\bf 1105} (2011) 027, arXiv:1101.2610
  [hep-ph].

\bibitem{Nojiri-gammaray-coulomb}
J.~Hisano, S.~Matsumoto, M.~M.~Nojiri, O.~Saito, \textit{Phys. Rev.} {\bf D71}
  (2005) 063528, hep-ph/0412403.

\bibitem{gunion-dmlight}
J.~F.~Gunion, D.~Hooper, B.~McElrath, {\it Phys.~Rev.} {\bf D73}, (2006)
  015011.

\bibitem{guniondm-10}
A.~Belikov, J.~F.~Gunion, D.~Hooper, T.~M.~P.~Tait, {\it Phys.
Lett.} {\bf B705} (2011) 82, arXiv:1009.0549 [hep-ph] \\
  J.~F.~Gunion, A.~Belikov, D.~Hooper, arXiv:1009.2555 [hep-ph].

\bibitem{daniel-astrolim}
D.~A.~Vazquez, G.~Belanger, C.~Boehm, arXiv:1107.1614 [hep-ph].

\bibitem{kaplinghat06}
L.~E.~Strigari, S.~Koushiappas, J.~S.~Bullock, M.~Kaplinghat, {\it Phys.~Rev.}
  {\bf D75} (2007) 083526, astro-ph/0611925.

\bibitem{cumberbatch11}
D.~T.~Cumberbatch, D.~E.~Lopez-Fogliani, L.~Roszkowski, R.~Ruiz de Austri,
  Yue-Lin~S.~Tsai, arXiv:1107.1604 [astro-ph.CO].

\bibitem{fermi-dsph}
A.~A.~Abdo {\it et.al}, {\it Astrophys.~J.} {\bf 712} (2010), arXiv:1001.4531
  [astro-ph.CO].

\bibitem{jaxodraw}
D.~Binosi, L.~Theussl, {\it Comput.~Phys.~Commun.} {\bf 161} (2004) 76,
hep-ph/030915.

\end{thebibliography}
\end{document}